\RequirePackage{fix-cm}

\documentclass{svproc}      
\usepackage{graphicx}
\usepackage{hyperref}
\usepackage[vertfit]{breakurl} 
\usepackage[]{footmisc}

\usepackage{url}

\begin{document}
\mainmatter  

\title{Decoding the Alphabet Soup of Degrees in the United States Postsecondary Education System Through Hybrid Method: Database and Text Mining}

\titlerunning{Hybrid Degree Level Prediction}

\author{Sahar Voghoei\inst{1}\inst{2}\inst{3}\and James Byars\inst{2} \and John A Miller\inst{1} \and Khaled Rasheedi\inst{1} \and Hamid A Arabnia\inst{1} }

\authorrunning{Sahar Voghoei et al.} 

\institute{Department of Computer Science,\\
\and Carl Vinson Institute of Government \\The University of Georgia, Athens, GA USA,\\
\and
\email{voghoei@uga.edu}
}

\maketitle              

\begin{abstract}
This paper proposes a model to predict the levels (e.g., Bachelor, Master, etc.) of postsecondary degree awards that have been ambiguously expressed in the student tracking reports of the National Student Clearinghouse (NSC). The model will be the hybrid of two modules. The first module interprets the relevant abbreviatory elements embedded in NSC reports by referring to a comprehensive database that we have made of nearly 950 abbreviations for degree titles used by American postsecondary educators. The second module is a combination of feature classification and text mining modeled with CNN-BiLSTM, which is preceded by several steps of heavy pre-processing. The model proposed in this paper was trained with four multi-label datasets of different grades of resolution and returned 97.83\% accuracy with the most sophisticated dataset. Such a thorough classification of degree levels will provide insights into the modeling patterns of student success and mobility. To date, such a classification strategy has not been attempted except using manual methods and simple text parsing logic.


\end{abstract}

\section{Introduction}

This paper develops a classification model whose task is to determine the degree levels (e.g., Bachelor, Master, Doctorate, etc.) earned by postsecondary students of American colleges and universities, observed in an unstructured national student-tracking dataset hosted by the National Student Clearinghouse (NSC). This model is intended to support further studies regarding student mobility (e.g., transferring institutions), which impacts educational success. While the model proposed in this paper is motivated by the demands of this area, many characteristics of the model address the challenges the nature of the data has raised to us. Section 1.1 below undertakes to expose the reader to our motives, while the challenges and problems will be described in section 1.2.

\subsection{Motivation}
Postsecondary educators have noticed that the increasing number of students who transfer across institutions has become a decisive factor in educational management. In 2019-2020, more than 1 million out of the 2.8 million students enrolled in the system fell in this category \cite{MediaCenter}. This rising trend is reshaping the traditional track of obtaining a college degree, where the concept of educational success has been historically defined with reference to students who begin a program at an institution and finish the same program in four years. This deviation distorts an institution's contribution to students' success because students who transfer are not considered. Adherence to the traditional reporting standards, for example, leads many schools to categorize those students that have transferred from their institutes with drop-out cases, while, since the cases of transfer do not necessarily and directly end in educational failure, this categorization system does not capture the changing nature of postsecondary education pathways. Moreover, in order to present a more successful picture of themselves, schools may be tempted to represent transferred students as non-transferred ones to cloud the institution's accountability in the student success pipeline. Likewise, schools with a high number of students who transfer-out but ultimately graduate elsewhere do not get the credit they deserve for their contribution to the student success pipeline. Misrepresentations of these sorts blur the educational as well as administrative data on which stakeholders rely.

The rate of student retention is one of the major indicators of institutional quality and prestige. If educators and administrators attain a clear and accurate understanding of factors that contribute to student transfer, they will be able to increase retention rates or navigate students through the most likely path towards educational success by the means of handling those factors. Such a handling, both on individual and institutional levels, will be more efficient if it is supported by the accurate predictions of the future stages of students' educational itinerary. Fore example, relevant treatments and resources can be efficiently channeled to student groups or individuals that, on the one hand, are predictably likely to transfer, and on the other hand, are predictably likely to achieve some extents of educational success. Such predictions cannot be made in absence of trustworthy and comprehensive data and precise analyses.

It is important to note that there are various patterns of transfer that can result in educational success or failure. These patterns and the motives behind them will be significant for educational policymakers whose objective is to optimize retention and graduation rates. For example, a report released in 2012 stated that 14.4\% of transfer students ended up completing a program shorter or a degree lower than what they had been admitted to at their initial schools (i.e., ‘reverse-transfer’) \cite{hossler2012transfer}. Therefore, in any effective report or forecast meant to help educators and policymakers, factors such as the level of pursued programs at both ends of a transfer track should be considered. 

\subsection{Problem}
One of the biggest steps towards addressing the aforementioned issues has been taken by the National Student Clearinghouse (NSC). This is a nonprofit organization that has partnered with nearly 3600 colleges and universities covering 97\% of all students in public and private U.S. institutions. NSC, collecting reports from its partners, facilitates tracking students by providing semester-based reports that indicate the institutions students have enrolled in, the standard length of the programs students have been admitted to, the first and the last date of attendance, the type of enrollment (e.g., full-time, part-time, withdrawn, etc.), and if the student has graduated in that semester. In the case of graduation, the degree title and the graduation date also will be given \cite{knuthwebsite}. All this data is based on the reports NSC receives from colleges and universities, and consequently, suffers from the typical deficiencies expected in this mode of data gathering; for example, features like majors are often missing or inconsistent. More importantly, this data offers little interpretative support to specify the types and directions of transfers. 

The shortcomings of the NSC student tracking data are most apparent when it comes to degree levels, which are indicated only by some segments embedded in the expressions for degree titles (e.g., in the degree title B.R.E., B denotes Bachelor). These expressions do not follow any straightforward standards. In fact, they appear in arbitrarily fashioned forms, including full words, short-hands, abbreviations, and any combination of the above forms so that, throughout the data, one degree title may be expressed in several forms. This heterogeneous aggregate of expressions is what NSC cumulatively inherited from the arbitrary styles of the reports composed by its partnering institutions. As a remedy, NSC issued a glossary that assigns each expression of a degree title in the data to a standard degree level, such as Bachelor or Master. NSC states that 70\% of these assignments have been manually verified based on inquires it had made with its partners, and the remaining 30\% was the result of a certain ‘text parsing logic’\cite{knuthwebsite}.A rule-based algorithm (conditional statements) comprises this ‘text parsing logic’.\footnote{NSC was generous enough to share with us the code that embodies this rule-based algorithm.}

NSC's strategy for classifying degrees suffers from a remarkable quantity of errors. Unfortunately, since we are not able to distinguish manual assignments from the ones generated by the rule-based algorithm,
.\footnote{NSC has not provided such a distinction.}
so far it is impossible for us to decide if one of the two methods is solely responsible for the errors, and consequently, should be revised. However, some of the rules are certainly responsible for some of the errors. For example, the rules classify all abbreviations that begin with ‘A’ as Associate Certificates, while there are cases in the data where AB stands for Art Bachelor. Errors of this sort render this glossary an unreliable source. Furthermore, a rule-based algorithm is not immune against unprecedented arbitrary expressions that will be introduced to the data in the future. In other words, in order to make sure a rule-based algorithm is able to handle new reports, the rules need to be manually updated after each occasion of data extension. All these concerns encouraged us to create a glossary of assignments, using recent text mining techniques. The remedy we are going to propose in this paper is a hybrid of a comprehensive database of the abbreviations of degree titles used by U.S. postsecondary educational institutions, and data mining and transfer learning methods that treat various expressions in the data. The aim of the current paper is to present our efforts, methods, results, and challenges with respect to this project.

\section{Background}

Several researchers have used NSC data to study the phenomenon of student transfer and its educational effects. Hale \cite{hale2019building} noticed the positive effect of transferring on the likelihood of program completion, which supports the capacities of transfer as a pathway for college and career success. Hale also explored the efforts that have been made to improve the quality of transfer experience with focus on research, policy, and practice. Umbach et al. \cite{umbach2019transfer} addressing the generally positive impact of the accessibility of community colleges and public universities on students’ success, stated that transferring to historically black colleges or universities is positively associated with GPA, college persistence, and degree completion \cite{voghoei2019university}. Some studies also used NSC data to account for the increase of the likelihood of graduation in the case of students who immediately transferred from two-year colleges to four-year programs \cite{nagaoka2020tracking}.

Although we didn't find any earlier work concerned with degree title abbreviations, there are papers that address abbreviations in text analyses. Several previous studies aimed to identify pairs of abbreviation-expansion or acronyms-definition in texts. They typically assumed that the pairs are typographically formatted, especially by putting the second part in parentheses (e.g., \cite{chang2002creating}). Larkey et al. \cite{larkey2000acrophile} described an ad hoc algorithm called Acrophile to extract acronyms from Web pages. Their approach is based on the use of parentheses and cue words. Schwartz et al. \cite{schwartz2002simple} reported a simple algorithm based on the use of parentheses and ad hoc rules for identifying the pairs of abbreviation-expansions in biomedical texts. It extracts short-form/long-form candidates from a text and then identifies the winning long-form for each short-form. Some other studies preferred heuristic methods. Wren et al. \cite{wren2002heuristics} developed a set of heuristics called Acronym Resolving General Heuristics (ARGH) to identify acronym-definition pairs, and Ao et al. \cite{ao2005alice} extracted abbreviations and their expansions from literature by the means of heuristic pattern-matching rules.

Hybrid approaches also have been adopted to deal with abbreviations. Typically, in such approaches a database will be manually or mechanically drawn from existing glossaries or sub-sections of a specific literature. In order to interpret abbreviations, this database will be consulted either prior or posterior to mining treatments if the latter is deemed necessary. Park et al. \cite{park2001hybrid} and Yu et al. \cite{yu2002mapping} are examples of this approach. The latter, whose interest is in genetic literature, have mined defined and undefined abbreviations in relevant texts. Then, in order to verify defined abbreviations, experts were consulted. For undefined abbreviations, on the other hand, labeling was constituted of mapping abbreviations to one of the four public databases, namely GenBank LocusLink, SWISSPROT, LRABR of the UMLS Specialist Lexicon, and BioABACUS. In genetic literature, where it is very likely to confuse nomenclatural codes for abbreviations, disambiguation appears as a serious challenge. In this respect, some effective solutions based on hybrid methods have been proposed by few researchers including Podowski et al. \cite{podowski2005suregene} and Dai et al. \cite{dai2010new}. 

The objective of our paper falls in the general category of short-text classification \cite{9422605}. Short texts are popular in today's web, especially with the emergence of social media. Inferring topics from massive amounts of short text has become a critical and challenging task for many content analysts. Conventional topic models, such as Latent Dirichlet Allocation (LDA)\cite{blei2003latent} and Probabilistic Latent Semantic Analysis (PLSA)\cite{hofmann2013probabilistic}, learn topics from document-level word co-occurrences by modeling each document as a mixture of topics. Cheng et al., addressing the deficiency of these methods in the face of the sparsity of co-occurrence patterns of short texts, proposed to learn topics by directly modeling the generation of word co-occurrence patterns in the corpus (Biterm Topic Models - BTM) \cite{cheng2014btm}. The limitations of contextual information, on the other hand, creates challenges in analysing short texts like Twitter messages \cite{voghoei2018deep}. Dos et al. \cite{dos2014deep} addressed this challenge by proposing a deep convolutional neural network (CharSCNN) that exploits from character-level to sentence-level information. The proposed network can explore the richness of word embedding produced by unsupervised pre-training proposed by Mikolov et al. \cite{mikolov2013efficient}, to which we are going to refer further in our research. In their sentiment analysis of short texts, Hassan et al. used CNN an LSTM for short text mining and focused more on local tensions than the dependency of sentences \cite{hassan2017deep}.

Pre-training word embedding models are also used in our project. We owe a good deal of our understanding of this concept to Mikolov et al., who displayed the capacities of word vectors through two major methods, namely Skip–Gram and Continuous Bag of Words (CBOW) \cite{mikolov2013efficient}. It was followed by Pennington et al. \cite{pennington2014glove} and Joulin et al. \cite{joulin2016fasttext}, who introduced two more popular word embedding methods. The former, being unsatisfied with the idea that the online scanning approach used by word2vec does not completely exploit the global statistical information related to word co-occurrences, proposed the concept of Global Vectors for Word Representation (GloVe), which is based on global matrix factorization, which applies the linear-algebraic process of matrix factorization to reduce large term frequency matrices representing the occurrences or the absence of words in a text. This method, called by its authors Latent Semantic Analysis (LSA), is used along with the earlier local context windows, namely CBOW and Skip–Gram \cite{pennington2014glove}. Joulin et al. proposed another word embedding model, namely FastText, wich is an extension of the word2vec model that represents each word as an n-gram of characters. It creates a word vector for a logically unrepresented word by the means of word's morphological characteristics. This helps to capture the meaning of shorter words and allows the embeddings to understand suffixes and prefixes, which are closer to abbreviations in terms of morphology \cite{joulin2016fasttext}.

\section{Data Structure and Challenges}

Out of the entire NSC data, we have extracted 6,972,614 rows, each representing information related with one student per one semester enrollment at a single educational institution in the University System of Georgia (USG). Among all features provided in this data, we are especially interested in the ones that are relevant to degree levels. These features are not limited to degree titles. For example, a correlation between degree levels and the standard duration of programs or the number of semesters within which students usually complete programs is fairly conceivable. However, degree titles are the features that are traditionally expected to contain elements directly corresponding to degree levels. 97,414 distinct expressions for degree titles are included in the entire NSC data. We will explain below that this is much more than the total number of distinct degree titles represented by these expressions. 

As mentioned earlier, the degree titles in NSC data are given in the form of abbreviations, short-hands, full-words, and any combination of these forms. In Table 1, a sample of the aforementioned expressions is given. The literature of text mining research teaches us that dealing with abbreviations and short-hands is complicated. In brief, two ways have been practiced to handle these categories: 1) considering these forms as `stop words' and removing them from the text \cite{silva2003importance}\cite{safadi2020curtailing}, which, due to the highly informative nature of abbreviations and short-hands in our data, is not what we need; 2) converting abbreviations and short-hands to their corresponding full-word expressions. To practice the latter way, as mentioned earlier, we will need to create a glossary in which each expression is assigned to a semantically independent interpretation in full-words. We could easily benefit from the previous work whose aim was to extract definitions and expansions from texts (as we reviewed in the background), only if for each assignment there existed a well formatted pair surfacing at least once in the data to associate full and partial expressions - a luxury this study does not have.
\begin{table}[ht]
\centering
    \includegraphics[width=8cm]{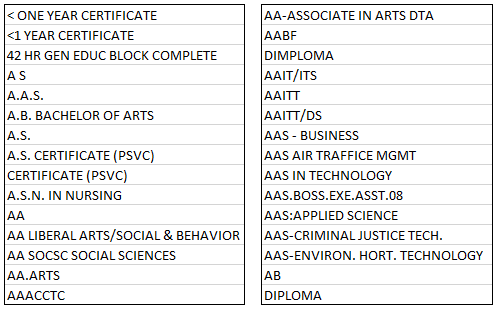}
    \caption{Sample of Degree Titles in NSC Data}
    \label{fig:DegreeTitle}
\end{table} 
The forms of expressions in NSC data posit more difficulties that specially challenge the detection of abbreviations and short-hands. For instance, it is difficult to distinguish short-hands from abbreviations and from misspellings. While in a normal text, abbreviations being all in uppercase are typographically distinct, in the data we are concerned with all texts are in uppercase, which neutralizes this distinctive feature of abbreviations.

The nature of data gathering also causes NSC data to suffer from lack of uniformity, so that one degree title, like Associate in Science, has surfaced in various forms in the following expressions: AS, A.S., A. S., *AS, A S, .A.S., A.S. CERTIFICATE (PSVC), A.S. IN NURSING, A.S.N. IN NURSING, AS.SCIENCE, ASN NURSING 2 YEAR RN, ASN TO BSN, etc., which defy any set of straightforward rules to identify them. We have noticed that in most of the expressions where two abbreviations are separated by a slash, the first component represents a degree level, while the second one is related with a field or a major (hence, the latter can be ignored or treated as a stop word because we are concerned with levels). However, there are cases like DTA/MRP in which none of the components directly provide information indicating a degree level (DTA stands for “Direct Transfer Agreement”, and MRP for “Major Related Program”), while the whole combination still represents an Associate degree in art and science. Moreover, there are expressions, like “MBA BBA”, “BACHELOR OF ARTS/MST”, or “BS AND MD”, that confusingly suggest more than one degree level. It will require us to employ more features, such as the standard duration of the program or the number of semesters within which students completed it, to label the degree level correctly. Adding to the chaos, there are expressions such as “BACHILLERATO EN ADM DE EMPRESA” or “BACHILLERATO EN ADMINISTRACION DE EMPRESAS”, which need to be saved from English spelling detection, lest they fall together with misspellings. The aforementioned examples, which represent a partial depiction of the challenging nature of the data, suggest that in addition to what text mining may contribute to our project, many more features should be introduced to the model.

\section{Method}

The objective of the present project is to propose a hybrid method to predict the degree level (e.g., Bachelor, Master, Doctorate, etc.) associated with every degree title expressed in NSC student-tracker data. Our method is an algorithm consisting in two major modules, namely database referral and Deep Learning models. With respect to the former, we created a comprehensive database of abbreviations that will be also used for evaluating our predictions. For Deep Learning, we developed a model composed of distinct components that work on text and extracted features classification separately. An overview of our algorithm can be given in the following three-step process:  
\begin{itemize}
  \item Step 1: Replacing abbreviations and short-hands with full expressions (see section 4.1.)
  \item Step 2: Creating a Deep Learning model to predict degree levels (see section 4.2-3.)
  \item Step 3: Evaluating and results (see section 5.) 
\end{itemize}
In the following sections, these steps will be discussed in details. 

\subsection{Abbreviations Replacement}
Usually, level related information embedded in full expressions of degree titles are more interpretable than those embedded in the corresponding abbreviations. It suggested to us to first replace abbreviations and short-hands in the data with their full expressions. However, the challenges we mentioned earlier had rendered the identification and replacement of abbreviations a complicated task. We created a database including as many abbreviations as possible officially used in the U.S. educational system to represent degree titles. In this database, each abbreviation is associated with a full expression as well as a degree level. This data has been extracted from authentic sources such as governmental websites and online college/university manuals. So far, the database covers 944 distinct abbreviations plus more than 10 short-hands. For every abbreviation in this database, we looked up in NSC data and replaced matching elements with the translations suggested by our database. 

\subsection{Data Pre-Processing}
Having replaced abbreviations and short-hands (see section 4.1), we have full-word expressions for most of degree titles. However, this text still suffers from spelling mistakes, irrelevant and insignificant words, a confusing variety of morphological cognates, etc. This requires a prepossessing that should include the following steps: 1) spell checking, 2) removing stop words, 3) stemming and lemmatization. These steps will be elaborated in the following sub-sections. 

\subsubsection{Spell Checking.}
In general, the methods of spell checking either work on lexical level (e.g, SymSpell and SpellChecker), or they work on contextual and syntactic level (e.g., Language-tool-python \cite{languagetool}, jamspell \cite{jamspell}, textblob \cite{loria2014textblob}, contextualSpellCheck \cite{yunus975context}). While the former only give spelling corrections, the latter can also pick grammatical errors. However, for our purpose, the latter are over sophisticated and too slow. 

Among the spell checking tools that work on lexical level, some use a distance/similarity algorithm, while some others benefit from modelings to calculate distance and probability. As the representatives of these two groups, we nominated pyspellchecker \cite{pyspellchecker} and SymSpellpy \cite{SymSpell}, respectively, to be tried against each other. SymSpellpy owes its nomination to its O(1) complexity maintained by the application of Symmetric Delete spelling correction algorithm and Hash Table. Furthermore, SymSpellpy lets us skip correcting phrases that match a given regular expression set, and also lets us set options to retain the case specifications (uppercase vs. lowercase) of abbreviations \cite{SymSpell}. In this way, we will be able to save the abbreviations that have survived the replacement (e.g., ANS or AC will not turn to AND or AS.) On the other hand, to find the most probably correct words, the other candidate, namely pyspellchecker, maps each word to an item on a word frequency list. It is based on a calculation in which the quantity of deletions, insertions, transpositions, alterations, and separations (permutations) that turn an incorrect word to a correct candidate will be involved \cite{mogotsi2010christopher}. 

\subsubsection{Stop Words.}
In many text mining operations, it is required to get rid of such words that do not contribute to the semantic level with which the mining is concerned. This class is commonly addressed as Stop Words. Words that appear with a high frequency (such as of, and, am, as, a, an, etc.) normally fall in this category. Studies have shown the highly positive effect of getting rid of Stop Words, even in the case of learning short texts \cite{siddiqui2021sarcasm}. However, in our project, a normal application of such a treatment will target the surviving abbreviations, which indeed, are vitally significant and informative. The remedy we adopt here is to periodically update the list of Stop Words before applying the method.  

\subsubsection{Stemming and Lemmatization.}
In texts such as the ones we are concerned with, as well as many other short texts, various morphological cognates bear the same semantics. It requires to reduce all such cognates to a single standard form or stem. This process, hence called Stemming, converts morphological variants, like “retrieval”, “retrieved”, “retrieves”, by stripping them of their affixes, to a common stem, like “retrieve”. Nevertheless, Lemmatization, which we adopted for our project, goes further and maps words to their dictionary forms (lemma). For example, while stemming may end up with “was" and “mice" as final stems, they can be lemmatised to “be" and “mouse", respectively.

We considered a number of approaches to lemmatization, among which many we found trading speed for a sort of sophistication that would not contribute to our needs. Some are able to deal with multiple languages (e.g., SpaCy \cite{spacy}), or mark syntactic roles (e.g., TreeTagger\cite{TreeTagger}) or dependencies and constituencies, sentiment, quote attributions, and relations (e.g., Standford CoreNLP\cite{CoreNLP}). Also, there are approaches like Gensim that are confined to a too limited number of lexical categories. Finally, for the current phase of our project, we adopted WordNet\cite{WordNet}, which is singled out for its speed. It is to say that we are also attracted to Standford CoreNLP for its performance on numbers, short texts, and dates (remember that there are date formats in the data.) This persuaded us to consider this method for our future work.

\subsubsection{Transfer Learning: Using Pre-trained Word Embeddings.}
We also approached transfer learning methods to use the knowledge attained by pre-trained word embedding, which captures the semantic and syntactic relations of words by training on large databases. The advantage of this approach has been already fairly established \cite{9346546}. There are two classes of embedding: character level and word level. In our project, we used GloVe \cite{pennington2014glove}, which is very close to Word2Vec \cite{mikolov2013efficient} and works on word level. GloVe focuses on words' co-occurrences over the whole corpus. Its embedding relates to the probability for two words to appear together. This makes GloVe faster and simpler than conceptualized embedding models such as ELMo \cite{peters2018deep}. We also used FastText \cite{mikolov2018advances}, which is a more recent method that maps a word to a vector composed of the sum of the n-gram segments (sub-words) into which the word has been broken. For example, the vector for the word “apple" can be the sum of the 3-gram sub-words “app", “ppl", and “ple". Since in our data many abbreviations appear as small segments embedded in a string of characters, we expected FastText to be capable of interpreting such problematic abbreviations.

\subsection{Classification Modeling}
Our model comprises two parts: 

1. Text classification (also known as text categorization) whose task is to assign labels (in our case: degree levels) to textual units (in our case: expressions for degree titles);

2. Feature classification that assigns the aforementioned labels to the sets of features or information related to students' records (in our case: the maximum standard duration of programs offered by an institute and the number of semesters within which a student completed the program). 

For the first part, we adapted a model of the class of CNN-BiLSTM \cite{yoon2017multi}. The second component of this combination, namely BiLSTM (Bidirectional-Long Short Term Memory) \cite{johnson2016supervised}, is an LSTM, which belongs to the wider class of RNNs (Recurrent Neural Networks). RNNs view a text as a sequence of words and aim to capture dependencies and structures\cite{shenavarmasouleh2020drdr}. However, LSTMs, devise memory to attend long term prior dependencies that basic RNNs would miss\cite{asali2020deepmsrf}. The advantage of BiLSTM is that it intends to capture dependencies on the both sides of a word. Since BiLSTM is a time-series deep learning model, it gives weight to sequence, while in the case of short texts, locality is important as well. Therefore, we balanced the BiLSTM with a model of the class of CNNs (Convolutional Neural Networks), which detects local and position-invariant patterns. What we have developed for our project was inspired by the high performance of Convolutional LSTM (C-LSTM) \cite{zhou2015c} and Dependency Sensitive CNN (DSCNN) \cite{zhang2016dependency}. 

\begin{figure}[hbt!]
\centering
    \includegraphics[width=12cm]{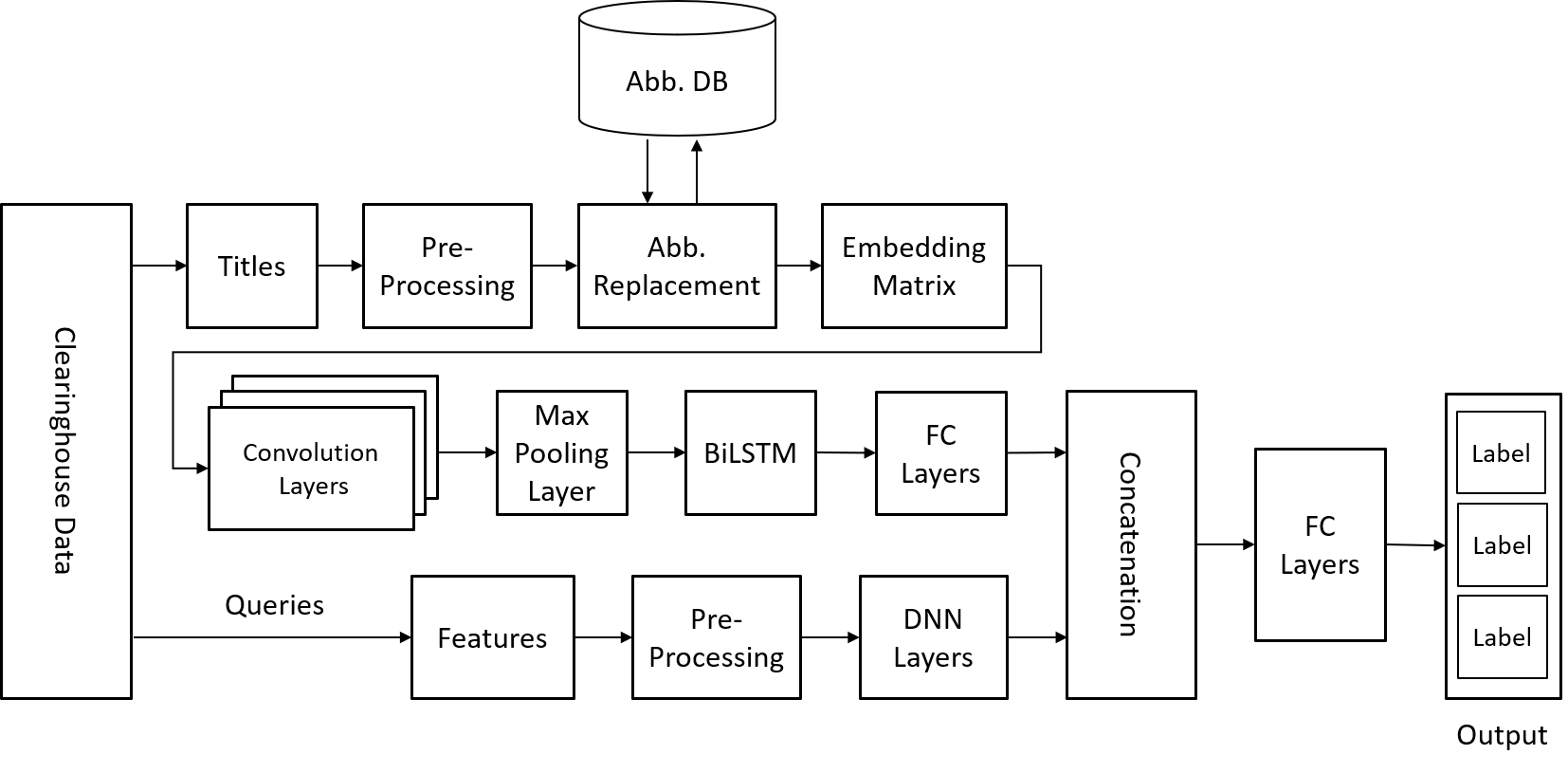}
    \caption{Overall Structure of the Algorithm}
    \label{fig:DLImage}
\end{figure}

In our model, the first layer of the text classification constructs an embedding matrix. Then, there will be convolutional layers followed by a max pooling layer. This, in its turn, will be followed by BiLSTMs and Fully Connected Neural Networks. Finally, after concatenating the latter's output and that of the feature classification (parallelly produced by prior layers of Deep Neural Network), there will be some layers of simple Fully Connected Neural Network that would label the samples. The structure of the entire algorithm is briefly illustrated in Figure \ref{fig:DLImage}.

\section{Result}

\begin{figure}[hbt!]
\centering
    \includegraphics[width=10cm]{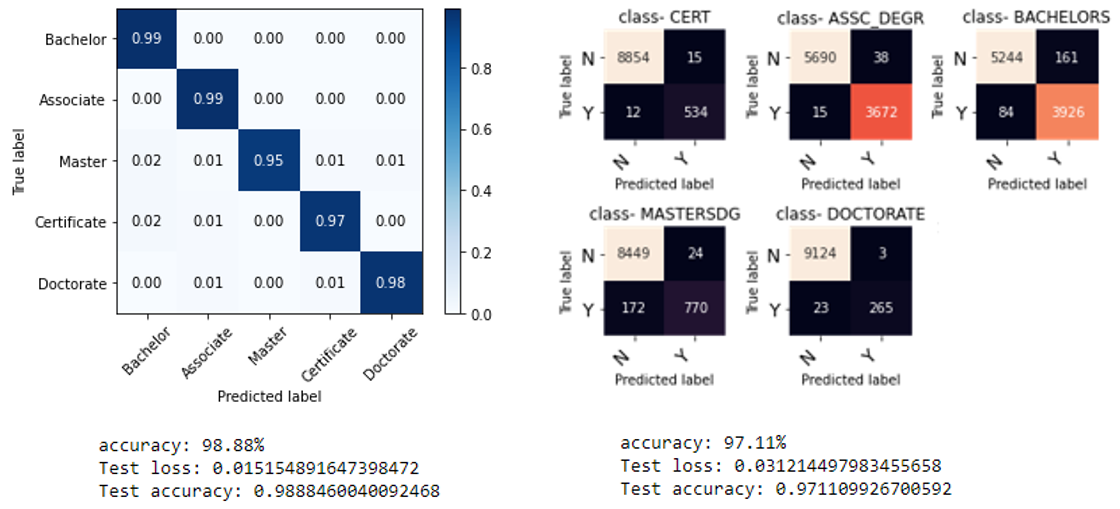}
    \caption{Confusion Matrices for Dataset 4: Multi-Class  (left) vs. Multi-Label (right)}
    \label{fig:confusionMatrix}
\end{figure}

Although NSC data has used seven labels (for their distribution see Table \ref{fig:DataStatistics}(B)), since the end our project is intended to serve demands a higher resolution, we adopted thirteen distinct labels (categories) for our model, in accordance with the degree levels of IPEDS (Integrated Postsecondary Education Data System). However, in order to adjust our model for different systems of degree evaluation, we designed four datasets graded from 1 to 4, such that for each dataset some categories of the prior dataset fall together to reduce the total number of categories. Table \ref{fig:DataSetsDefinitions} shows the categories of the four datasets across which our thirteen original categories gradually collapse to five categories.  
\begin{table}[hbt!]
\centering
    \includegraphics[width=\textwidth]{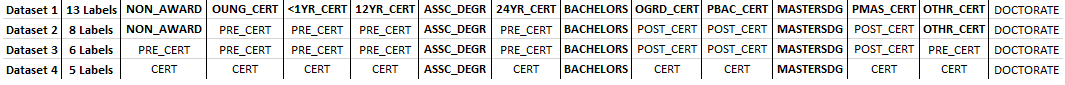}
    \caption{Correspondent Categories of 4 Graded Datasets}
    \label{fig:DataSetsDefinitions}
\end{table}
The total number of unique samples in NSC data is 97414, out of which 3382 have been labeled as “missing”. It is to mention that after we relabeled the samples, we realized that 94\% of “missing” labels are in fact of Doctorate level. Having referred to experts, we re-categorized 48508 samples under the thirteen categories, whose detailed distribution is given in Table \ref{fig:DataStatistics}(A), where samples that are assigned to more than one label are counted repeatedly. 
\begin{table}[hbt!]
\centering
    \includegraphics[width=10.5cm]{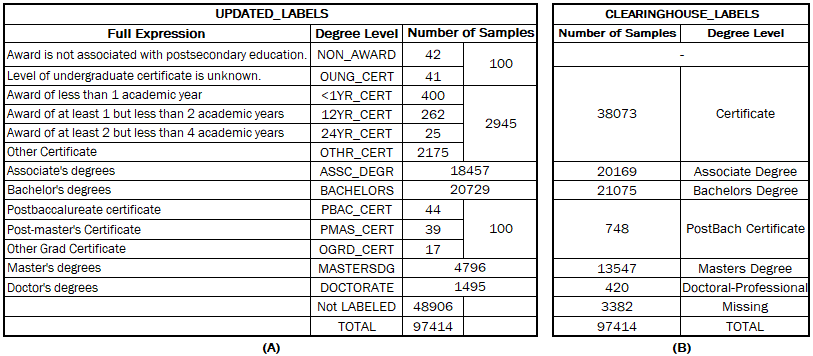}
    \caption{Distribution of Updated Labels (A) vs. NSC Labels (B)}
    \label{fig:DataStatistics}
\end{table} 
In terms of classification, NSC dataset is Multi-Class, i.e., it assigns each sample to a single category out of many. However, we noticed at least 302 samples that need to be labeled with more than one category. For example, MASTER OF BUSINESS ADMINISTRATION/DOCTOR OF PHARMACY should be labeled as Master as well as Doctorate (more examples can be found in Table \ref{fig:DegreeLavelDatabase}). Examples of this sort imply that labels can be realistically assigned only by the means of a Multi-Label classification that assigns each sample to a set of labels. Naturally, this commitment to realism is expected to cost the accuracy of the model. To assess this cost, we trained our model with a Multi-Class version of dataset 4 in which a multi-labeled sample is mapped into the highest level within its label set (Figure \ref{fig:confusionMatrix}-right). On the other hand, Figure \ref{fig:confusionMatrix}-left shows our results with the Multi-Label version of the same dataset. As expected, the former predicts labels with 98.88\% accuracy, while the latter's accuracy is 97.11\%. However, from now on, our experiments will be limited to Multi-Label classifications.  
\begin{table}[ht]
\centering
    \includegraphics[width=10cm]{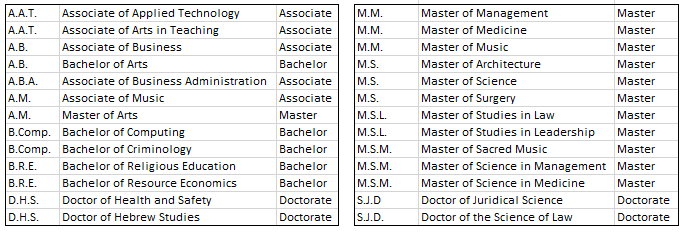}
    \caption{Sample of Multi-Assignments in Database of Abbreviations}
    \label{fig:DegreeLavelDatabase}
\end{table} 
We ran some experiments to decide about the combination of external packages we would employ for the pre-processing steps and embedding. In all those experiments, we used WordNet for Lemmatization. For the choice of spell-check we ran the combination of abbreviation replacement and the embedding method of GloVe, alternatively with SymSpellpy and pyspellchecker, of which the results are shown in columns III and IV of Table \ref{fig:AccuracyTable}, respectively. The surpassing results of pyspellchecker fixed it in our pre-processing. Then, to verify the contribution of abbreviation replacements, we ran our model again with GloVe and pyspellchecker but without an abbreviation replacement, which performed with a remarkably lower accuracy for all datasets (see Table \ref{fig:AccuracyTable}-column II). After fixing the combination of abbreviation replacement and pyspellchecker, we replaced GloVe with FastText, which slightly improved the accuracy (see Table \ref{fig:AccuracyTable}-column V). All these results were generated with 10 Cross Validation. Table \ref{fig:recall} shows the accuracy of our text mining model with the finalized combination of external packages for each category in dataset 1 (the highest label resolution). Here, the lowest performance is associated with the categories of “Certificates”. This explains why, as Column V in Table \ref{fig:AccuracyTable} indicates, the collapse of the categories of Certificates together across database 1 to 4 increases the overall accuracy of the model.      
\begin{table}[hbt!]
\centering
    \includegraphics[width=10.5cm]{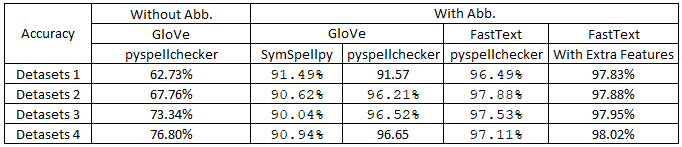}
    \caption{Accuracy of the Model with Different Sets of External Packages}
    \label{fig:AccuracyTable}
\end{table}
\begin{table}[hbt!]
\centering
    \includegraphics[width=5.5cm]{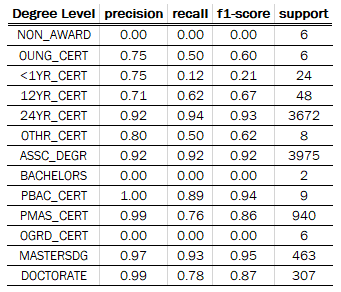}
    \caption{Accuracy of the Model by Categories}
    \label{fig:recall}
\end{table}
We extracted two features from the NSC data, namely the maximum standard duration of programs offered by a school and the number of semesters within which a student completed a program. We crossed the NSC dataset with our four graded datasets based on degree titles as common features. Column VI in Table \ref{fig:AccuracyTable} shows the improvement of results for each dataset after adding these features to the model. In fact, this column represents the final performance of our model, which predicts the labels of our most sophisticated dataset (13 categories) with 97.83\% accuracy. However, the accuracy rises up to 98.02\% in the case of the least sophisticated dataset (5 categories).

\section{Future Work}
We have planned a set of quantitative improvements to get better results in the future. This will include expanding the database of abbreviations and increasing the number of labeled samples for the purpose of training (so far we have 58906 unlabeled samples). Currently, the distribution of labels is biased towards Associate and Bachelor levels, which dominate the data. In order to address this unbalance, our quantitative improvements will target minority categories. In addition, we are considering GANs (Generative Adversarial Networks) for this purpose. With respect to preprocessing, since the number of Spanish expressions in the data is remarkable, we have planned to extend the spell checking to Spanish as well. We also need to periodically seek experts' opinions in the case of incorrect predictions, as well as selective sets of newly labeled samples, so that corrected and approved labels can be added to the training set. As for transfer learning, we will try more embedding packages and models, such as character level learning models. The challenge raised by multi-labeled samples that are simultaneously associated with several distinct majors or levels inspired us to consider the possibility of labeling on student level instead of title level. 


\section{Conclusion}
The rising quantity of transfer students and the variety of the patterns of transferring deserve to be studied as impactful contributors to educational success. This requires the precise data of the academic levels (e.g., Bachelor, Master, Doctorate, etc.) through which students transfer across the postsecondary educational system. Although the National Student Clearinghouse (NSC) has undertaken making such reports accessible, the inconsistencies mostly imposed by the nature of its methods of data gathering make its reports partly un-interpretable. To address this problem, we developed a hybrid method whose task is to predict the academic levels of the mostly nonstandard expressions for degree titles given by the NSC student-tracking data. This hybrid method comprises two modules. The first module replaces abbreviations and short-hands with their full expressions that include degree levels. This will be done by looking up in a comprehensive database including nearly 950 abbreviations officially standing for degree titles, which we extracted from authentic documents and manuals issued by American institutes of postsecondary education and related organizations. The second module is a combination of feature classification and text classification modeled with CNN-BiLSTM. The training of this model was preceded by several steps of heavy preprocessing. For each step, we considered several methods among which we tried the best candidates and chose the ones that proved to be appt for our data and our objective. We also compensated the limits of the size of the data with the prior knowledge provided by an embedding model. We ran several experiments with different scenarios for pre-processing and embedding that finally, in addition to verifying the contribution of the abbreviation replacement, led us to SpellCheker, WordNet, and FastText as the external packages to employ in our project. Furthermore, the purpose of our model required to train it with four Multi-Label datasets of different numbers of categories. The accuracy of our predictions for each of these datasets sorted by resolution is: 97.83\% (for 13 categories), 97.88\% (for 8 categories), 97.95\% (for 6 categories), and 98.02\% (for 5 categories). 

\nocite{*}
\bibliographystyle{spmpsci}
\bibliography{main}

\end{document}